\documentclass[12pt]{iopart}

\pdfoutput=1
\usepackage{hyperref}
\usepackage{cite}
\usepackage{graphicx}
\begin{document}

\title[High energy cosmic ray induced atmospheric ionization]{Lookup tables to compute high energy cosmic ray induced atmospheric ionization and changes in atmospheric chemistry.}

\author{Dimitra Atri$^{1}$, Adrian L. Melott$^{1}$ and Brian C. Thomas$^{2}$.}

\address{$^{1}$University of Kansas, Department of Physics and Astronomy, 1251 Wescoe Dr. \# 1082, Lawrence, KS 66045-7582, United States of America}
\address{$^{2}$Washburn University, Department of Physics and Astronomy, Topeka, KS 66621, United States of America}

\ead{melott@ku.edu}
\begin{abstract}
A variety of events such as gamma-ray bursts and supernovae may expose the Earth to an increased flux of high-energy cosmic rays, with potentially important effects on the biosphere. Existing atmospheric chemistry software does not have the capability of incorporating the effects of substantial cosmic ray flux above 10 $GeV$. An atmospheric code, the NASA-Goddard Space Flight Center two-dimensional (latitude, altitude) time-dependent atmospheric model (NGSFC), is used to study atmospheric chemistry changes. Using CORSIKA, we have created tables that can be used to compute high energy cosmic ray (10 $GeV$ - 1 $PeV$) induced atmospheric ionization and also, with the use of the NGSFC code, can be used to simulate the resulting atmospheric chemistry changes. We discuss the tables, their uses, weaknesses, and strengths.

\end{abstract}

\vspace{2pc}
\maketitle

\section{Introduction}
Nearby supernovae [1], gamma ray bursts [2, 3] and possibly galactic shocks [4] may bathe the Earth in cosmic rays (CRs) of much higher than usual incident energies. It is of considerable interest to investigate the effect of such events on the Earth's atmosphere, and consequent possible connections to mass extinctions and other events in the fossil record.  Until now, cosmic rays of this high energy have been included only by means of a simple phenomenological approximation, enhancing the existing background CR ionization, or not at all.  Computations of the effects of gamma-ray bursts [2] have been based on only the effects of photons.  However, a sufficiently strong background of very high-energy cosmic rays can punch through the galactic and terrestrial magnetic fields [3] and irradiate the Earth, with effects potentially competitive with those of photons.  Other potential scenarios which may account for a 62 My periodicity in biodiversity [4] are based solely on CR, and to date the terrestrial effects have only been approximated.  Thus there is considerable value in developing software to model the effects of a spectrum of CR with energies above those normally included in atmospheric computations.

We have developed a method to model changes in atmospheric chemistry when high energy cosmic rays (HECRs) ionize the atmosphere.   When HECRs hit the atmosphere, like other CRs they will interact with atmospheric constituents, primarily the molecules of $N_{2}$ and $O_{2}$. This interaction will either result in a nuclear reaction or electromagnetic interaction ionizing the atmosphere, the latter being the interaction of primary importance for chemistry [5]. To study this interaction we created a normalized HECR ionization spectrum data table that contains ionization energy due to HECRs at various energies and altitudes.  We will discuss the methods used, how the data table was generated, and its use in a widely used time-dependent atmospheric ionization and chemistry code.  This result naturally complements and extends basic work on the lower-energy CRs that normally dominate atmospheric ionization [6].

In order to obtain ionization energy from HECRs, we used CORSIKA (COsmic Ray SImulations for KAscade), a high energy cosmic ray extensive air shower simulator [7, 8]. The code follows the interactions of a primary cosmic ray and its secondary particles through the atmosphere to the ground.  CORSIKA uses Monte Carlo calculations to account for high energy strong and electromagnetic interactions using a number of extensive air shower simulators (of which we used UrQMD-low energy and EPOS-high energy) [9].

Using CORSIKA we created tables so that atmospheric ionization can be calculated for an arbitrary CR spectrum between 10 $GeV$ and 1 $PeV$.  We intend to continuously update this table to higher energies, eventually reaching the highest energies observed.

\section{Methods}
The tables we have generated give energy deposition binned in two different ways, and can be used to either simply compute atmospheric ionization or can be used with the NASA-Goddard two-dimensional atmospheric chemistry code, hereafter NGSFC.

The NGSFC code has been used extensively to study the effects of solar proton events on the atmosphere [10]. It has also been used to study supernovae [11], solar flares [13] and gamma-ray burst effects [2].  We will only briefly describe this code, given detailed accounts elsewhere [2, 12, 14].  There are 58 log pressure bands (we use only the first 46 of these bins here, as discussed below) and 18 bands of latitude.  The model computes atmospheric constituents with a largely empirical background of solar radiation variations, with photodissociation, and including small scale mixing and winds.  Also, it includes an empirical background of CR source ionization based on current levels, which includes an 11-year solar modulation cycle, all with a one day timestep.

\begin{figure}[htp]
\centering%
\includegraphics[scale=0.7]{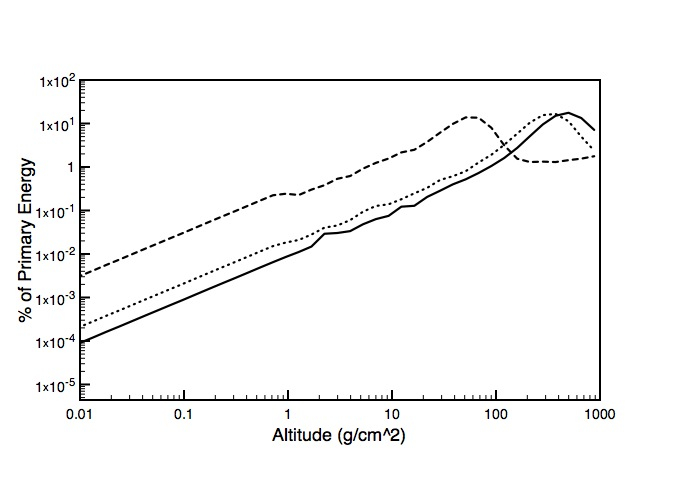}
\caption{Fractional energy deposition for a 10 TeV primary at zenith angles $5^{o}$ (solid), $45^{o}$ (dotted), and $85^{o}$(dashed) in bins of the NGSFC code in log pressure, proportional to total column density traversed by the shower, which is approximately linear in altitude.}
\label{figure 1}
\end{figure}

CRs of $10^{10} - 10^{15}$  $eV$ are of much larger primary energy than those that dominate normal galactic CRs, so one should not simply turn up the usual background, as in a previous supernova study [11]. We used CORSIKA, which is designed to perform detailed simulations of extensive air showers initiated by high energy cosmic ray particles.  We did 25 simulated showers at each of a series of energies by 0.1 in log$_{10}$ intervals of primary energy between 10 $GeV$ and 1 $PeV$, i.e. at 50 different primary energies.  

We created lookup tables using data from CORSIKA runs that contains atmospheric ionization energy deposition per primary due to CRs in the range of 10 $GeV$ - 1 $PeV$.  The sum of deposition over altitude is less than the total of the primary energy, as not all energy is deposited in the the atmosphere by electromagnetic processes. Nuclear interactions also occur between HECRs and atmospheric particles, but nuclear energy is dumped into nuclei, mostly not into atmospheric chemistry. Energy that goes into nuclear interactions or reaches the ground is not included in the deposition table.

An arbitrary spectrum can be convolved with this data and the results used in the NGSFC code (or other similar codes) to simulate the effects this energy flux deposition will have on atmospheric chemistry. We investigated the effect of zenith angle by running 100 shower ensembles (1500 primary particle simulation runs) at zenith angles of $5^{o}$, $45^{o}$, and $85^{o}$ at $10^{13}$ $ eV$. In Figure 1 we show the fractional energy deposition for each of these zenith angles (excluding nuclear interactions) per interval (intervals of the NGSFC code) in log pressure, proportional to total column density traversed by the shower, which is approximately linear in altitude.  Note that the lateral displacement in the lines, and the location of their maxima, are reasonably approximated by a $(cos$ $ \theta)^{-1}$ factor, confirming the simplest thing one would expect from a column density factor.  For a general-purpose code at the energies we consider, it is appropriate to assume the flux is isotropic. For each shower, we recorded the fractional energy deposition in each of 1000 bins of 1 $g$ $cm^{-2}$  column density. We used their mean to construct a lookup table describing the energy deposition for a particle of given primary energy as a function of pressure.  Below we describe in detail the construction and use of the table.

The greatest deposition of energy per bin corresponds to the first interaction between the incoming CR and the atmosphere, which occurs very high in the atmosphere.  Since log column density is nearly linear with altitude, we analyze the maximum energy deposition per bin of log column density.  This has weak trends with primary energy, as shown in Figure 2. 
This gives about 400 $g$ $cm^{-2}$ as the site of the greatest deposition of energy per unit distance, corresponding to an altitude of about 5 $km$. This can be compared with (a) about 13 $km$ as the mean altitude of maximal energy deposition density for the normal CR spectrum as implemented in the NGSFC code, strongly biased toward latitudes greater than above about 60 degrees, and (b) 22 to 35 $km$ as the peak deposition for $keV-MeV$ photons, depending upon energy [2,15]. 

The table was originally created to compute ozone ($O_{3}$) depletion and other atmospheric chemistry changes [5] resulting from energy deposition by HECRs.  $O_{3}$ lives at altitudes of 10-35 $km$ [16] with considerable latitude dependence [15]. Our data is inaccurate from 46-90 $km$ because CORSIKA runs on a linear column density scale starting with 1 $g$ $cm^{-2}$ , but main effects of atmospheric chemistry changes on the biosphere occur at lower altitudes. A 46-90 $km$ altitude is less significant because high-energy CRs rarely have their first interaction that high in the atmosphere.

\begin{figure}[htp]
\centering%
\includegraphics[scale=0.7]{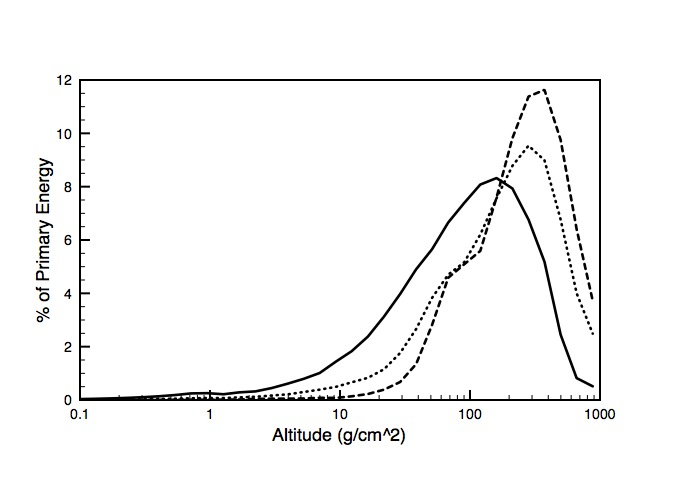}
\caption{Electromagnetic energy deposition from a 100 $GeV$ primary (solid), a 10 $TeV$ primary (dotted) and an $PeV$ primary (dashed) per $g$ $cm^{-2}$, in bins of NGSFC code . Energy deposition for the higher energy primaries is deeper in the atmosphere.  Energy going into nuclear interactions or hitting the ground is not included in this table}
\label{figure 2}
\end{figure}
\section{Construction of the Lookup tables}
The CORSIKA model 6.900 is used in all simulations. When installing CORSIKA, UrQMD was used for the low energy hadronic interaction model and EPOS as the high energy hadronic interaction model [22]. CORSIKA was installed with CURVED, SLANT and UPWARD options to simulate primaries at large zenith angles and track upward going particles in a curved atmosphere. The following are the important variables used in the input file for CORSIKA. The Longitudinal Shower Development variable (LONGI)  is set as LONGI T 1 T F giving longitudinal ionization for every 1 $g$ $cm^{-2}$ bin to facilitate interpolation to the bins of GSFC photochemical code. The variable THETAP gives the angle of incidence of particles and was set as THETAP z z, where z is the zenith angle. The code ran for zenith angles $5^{o}$, $15^{o}$...upto $85^{o}$, a total of 9 angles. The energy range variable (ERANGE) is set as ERANGE x x, where x is the variable energy of the primary particle. For our preliminary runs (10 $GeV$ - 100 $TeV$), CORSIKA was run 100 times at each of the energies stated above using different SEED random number variables from the input file for each run. The data file outputs a variety of information. The sum of the average longitudinal energy deposited in GeV given at every 1 $g$ $cm^{-2}$ is the only data we are interested in order to evaluate the atmospheric ionization. In the future we plan to extend this table above the $PeV$ range.

	CORSIKA ran at 51 different energy levels 100 times each at intervals of 0.1 in log energy for 9 angles. Each time it was run CORSIKA simulated 15 protons entering the atmosphere. Therefore, we compiled 13500 particles for each of the 51 energy levels. The energy was averaged for each of the 900 runs at each longitudinal pressure bin. 
	
		Minor problems arose when transferring the CORSIKA data to the NGSFC code. First, the smallest atmospheric depth bin at which energy deposition is given is 1 $g$ $cm^{-2}$ in CORSIKA, while 22 of the 46 pressure levels are below 1 $g$ $cm^{-2}$ in the NGSFC code. Secondly, CORSIKA outputs deposition in a linear pressure scale. We interpolated to a logarithmic scale for the NGSFC code. We converted CORSIKA atmospheric depth units of $g$ $cm^{-2}$ to millibars (needed for NGSFC) by a conversion factor of 0.98.

		To resolve the problem of CORSIKA's highest interaction altitude being below 22 of the NGSFC codes altitudes we linearly interpolated the data using the ionization from the 1 $g$ $cm^{-2}$ bin. Because the first interaction point of primaries is rarely within the first $g$ $cm^{-2}$, most of the CORSIKA output files have no energy deposition for this altitude. Also, the data we used from CORSIKA is binned in 0.1 $GeV$ intervals, meaning for the lower energy runs many of the 1036 altitude levels are zero.  For this reason our output file may have no energy in the 22 highest altitude bins for certain energy levels. For our purpose of looking at the ozone these higher altitudes, above $\sim$ 46 $km$, are not important, since the amount of ionization in them from HECR would be quite small.

		We use the same linear interpolation method used for each bin with less than 1 $g$ $cm^{-2}$, depositing the energy from ground level to the highest logarithmic altitude level in the last bin. The logarithmic altitude bins are centered on the bins used in the NGSFC code (Table 2).
		
		The geomagnetic field should be mentioned with regard to the lookup table. Its impact on CRs is minor for protons with energies greater than 17 $GeV$ [17]. Because this only impacts a small region of the CR range of importance (10 $GeV$ - 1 $EeV$) effects due to the magnetic field are small.  If desired, alterations to the latitude distribution of energy deposition can be made in the 10 to 17 $GeV$ range to simulate the effects of deflection by the geomagnetic field.

\section{Comparison with existing simulations}
We have compared our data with existing data generated by CORSIKA [6] as well as analytical methods proposed by Velinov et al. (2008) [21].

We found very good agreement with Usoskin et al. data at 1 $TeV$ (figure 3), 100 $GeV$  (figure 4) and 10 $GeV$ (figure 5). However, there is about 5$\%$ higher ionization in our data at lower altitudes. This can be attributed to the UrQMD model we have used, which is more efficient in tracking muons compared to FLUKA used in the earlier work. Since muons are a major source of atmospheric ionization, especially near the ground, we would expect this enhancement. It must also be noted that we used the latest version of the code (CORSIKA 6.900, 2009) compared to CORSIKA 6.204 (2005) used in the earlier work. 

We also compared the shower profile at 1 $PeV$ (figure 6) with Tanguy Pierog's (CORSIKA developer) simulation generated by the same hadronic interaction models and found very good agreement. This simulation provides the energy deposition profile of a 1 PeV primary using the EPOS-UrQMD hadronic interaction models produced with CORSIKA. Fluctuations in the energy deposition profile were removed using the CONEX code. 

The analytical method [21] can be used for a thin target, which is above 50 km altitude. Also, the equation used to calculate ionization limits primary energy range of 600 $MeV$ Ð 5 $TeV$. We applied the analytical solution to the energy range 17 $GeV$ Ð 5 $TeV$ in order to ignore the effects of solar modulation. Our values of ionization are about $5\%$ higher than predicted by the approximation (Table 3). As discussed earlier,  CORSIKA does not allow altitude bins smaller than 1 $g$ $cm^{-2}$, which contains the entire atmosphere above 50 km, and therefore, it is not possible to resolve different levels within 1 $g$ $cm^{-2}$ bin. 

Atmospheric ionization produced by the primaries in our energy range is a very tiny fraction of the total ionization from the GCR flux, and therefore we can not compare our results with certain models [20], but it will be possible to compare for a given CR spectrum.

\newpage
\begin{figure}[htp]
\centering%
\includegraphics[scale=0.7]{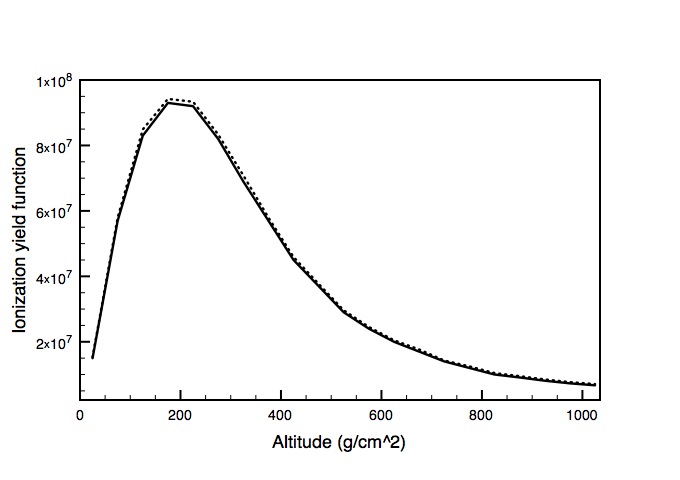}
\caption{Comparison of our data (dash) at 1 TeV with Usoskin et al. (solid) [6]. }
\label{figure 3}
\end{figure}

\newpage
\begin{figure}[htp]
\centering%
\includegraphics[scale=0.7]{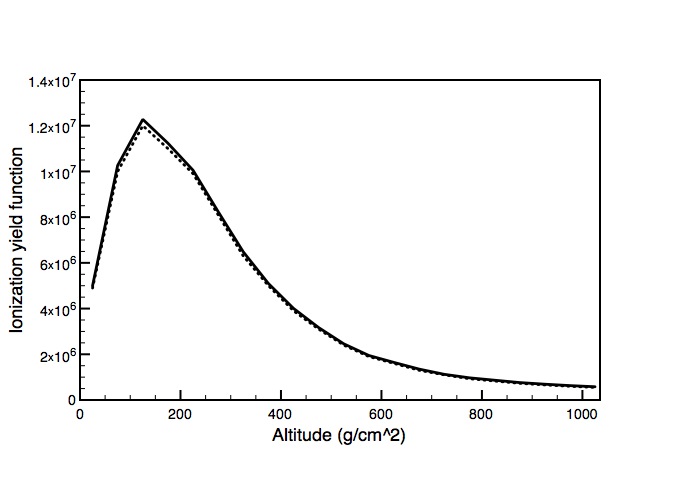}
\caption{Comparison of our data (solid) at 100 GeV with Usoskin et al. (dash) [6]}
\label{figure 4}
\end{figure}

\newpage
\begin{figure}[htp]
\centering%
\includegraphics[scale=0.7]{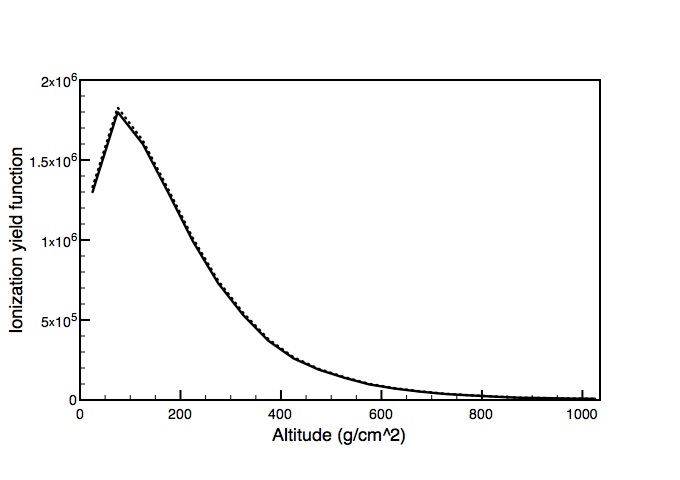}
\caption{Comparison of our data (dash) at 10 GeV with Usoskin et al. (solid) [6]}
\label{figure 5}
\end{figure}

\newpage
\begin{figure}[htp]
\centering%
\includegraphics[scale=0.7]{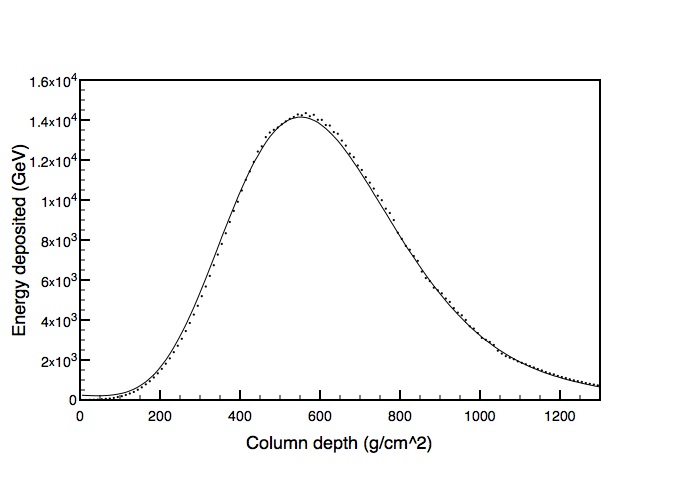}
\caption{Comparison of our data (dots) at 1 PeV with T. Pierog's data (solid) produced with CORSIKA/CONEX in bins of 10 $g$ $cm^{-2}$}
\label{figure 5}
\end{figure}

\section{How to use the lookup tables}

The lookup tables are formatted into 51 columns corresponding to 51 energies but the number of rows for both tables are different. For the atmospheric  chemistry model, the table has 46 rows corresponding to the altitude bins of the NGSFC code shown in table 2. As is, it displays ionization energy deposition for a spectrum of 1 particle per steradian per square meter per second for every 0.1 logarithmic energy bin, in units of $GeV/m^{2}.sr.s$. Trivially, one must multiply by the number of particles per unit area per steradian per second in their bin to get the total energy flux at each altitude bin for a given spectral form. To input this data into the NGSFC code the energy at each altitude must be added over all energy levels, creating a 1-dimensional data set of total energy deposition flux at each altitude. 
	
The general purpose table has 51 columns as mentioned above and 21 rows of bin size 50 $g$ $cm^{-2}$ each. The table displays the ionization yield function in units of ion pairs per meter square per steradian per second. Total number of ion pairs for a given spectrum can be obtained by the method mentioned above.

As mentioned previously, the NGSFC code takes ionization energy flux in a 2-dimensional format using altitude and latitude. The data is now in 1-dimension with respect to altitude, so the user must create a latitude component. The NGSFC code has 18 latitude bins ($90^{o}-80^{o}, 80^{o}-70^{o}, 70^{o}-60^{o} $ etc). For an isotropic flux, the same flux is entered for each latitude bin. The final data file to be input into the NGSFC code should now be a 2-dimensional set of ionization energy flux deposition at the 18 latitude bins for 46 altitudes, a total of 828 data points.  For a point source, the input into latitude bins may be adjusted by the appropriate factor, including a correction for the $cos$ $\theta$ factor as in [2].  This may result in hemispheric differences in the results.

The energy flux data as a function of altitude and latitude, generated as described above, may be used in the NGSFC code by way of a simple read-in subroutine. Depending on the units of the spectrum used, conversion to $cm^{-2} s^{-1}$ may be necessary since the NGSFC code uses $cgs$ units.  In order to use the input as a source of $NO_{y}$, the energy deposition rate must first be converted to ionization rate. This is accomplished using 35 $eV$ per ion pair [18], which finally gives values in units of ions $cm^{-2} s^{-1}$.  Constituents in the code are stored with units of number $cm^{-3} s^{-1}$.  Therefore, the area ionization rate is converted to a volume rate using the height of the altitude bins, which depends on the current density of each bin at read in (the density depends on temperature, which depends on the presence of sunlight, etc.).  This ionization rate is then used as a source for $NO_{y}$, assuming 1.25 molecules are created for each ion pair [16].  The model then runs as usual, incorporating this source of $NO_{y}$ in the relevant chemistry computations.  The general procedure here is the same as that used in previous work with both photon and solar proton ionization sources [2, 10].

It must be noted that alpha particles were not considered to produce these tables. At such high energies, an alpha particle can be approximated by four protons with the same energy per nucleon and the resulting ionization can be computed using the data generated with protons.

\section{Discussion}

The inclusion of CR effects will make possible a more accurate investigation of the effects of gamma-ray bursts on the Earth, which have the potential to explain certain mass extinction events [19].  A quantitative treatment of possible time-varying flux of high-energy CRs becomes possible [5]. It may have other applications which we cannot anticipate, and should be made generally available. 

This lookup table will be made freely available via ftp, and upgraded in the future as appropriate.  Look for a link at $http://kusmos.phsx.ku.edu/~melott/Astrobiology.htm$ upon publication of this paper.

\ack

We thank Tanguy Pierog  and Dieter Heck for their help and advice in using CORSIKA. This research was supported in part by the National Science Foundation through TeraGrid resources provided by the National Center for Supercomputing Applications and by NASA grant NNX09AM85G. B.T. acknowledges a Small Research Grant from Washburn University.

\newpage
\begin{table}
\centering%
\includegraphics[scale=0.5]{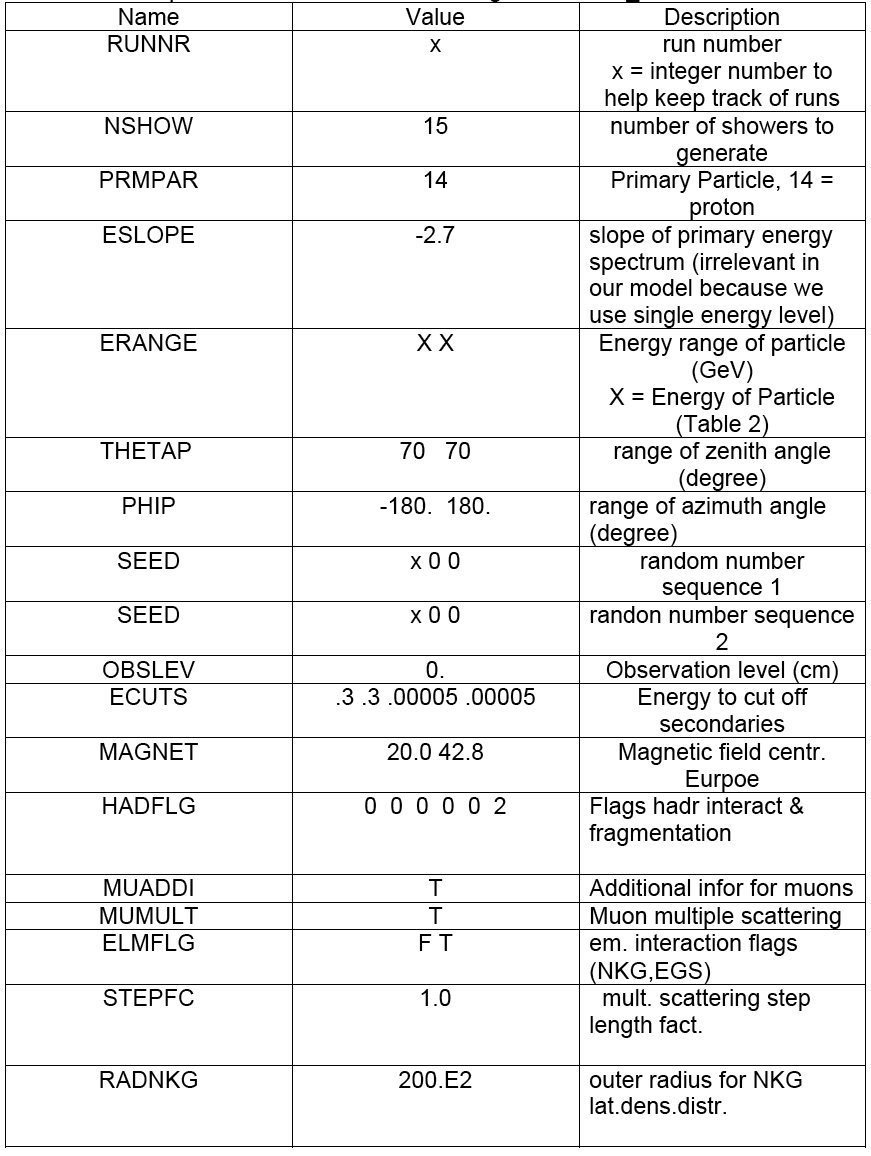}
\end{table}
\begin{table}
\includegraphics[scale=0.5]{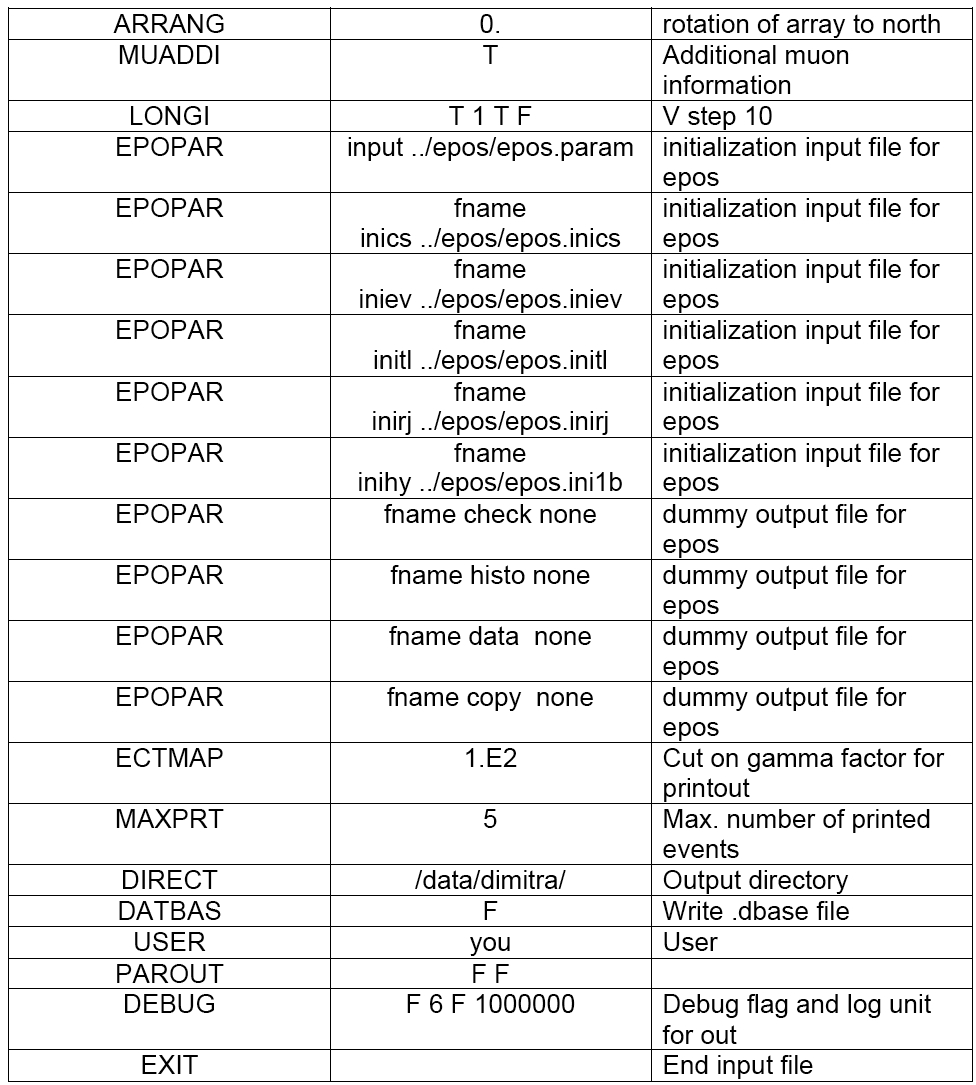}
\caption{Input File for CORSIKA - Complete Description of Variables at $http://www-ik.fzk.de/corsikausersguide/corsika_tech.html$}
\end{table}

\newpage
\begin{table}
\centering%
\includegraphics[scale=0.45]{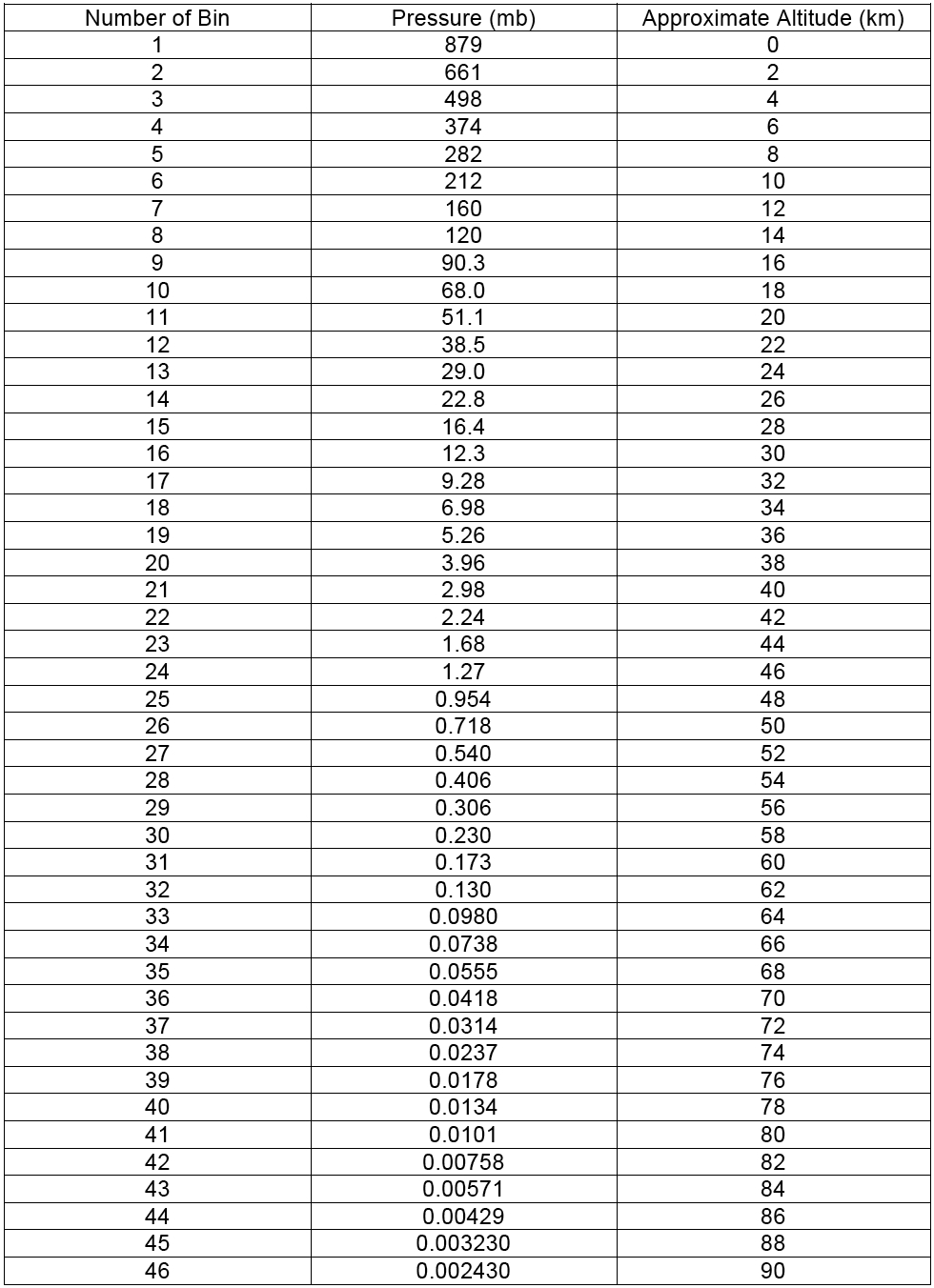}
\caption{Altitudes for NGSFC code where pressure values correspond to the center and altitude values correspond to the lower bound of the bin.}
\end{table}

\newpage
\begin{table}
\centering%
\includegraphics[scale=0.45]{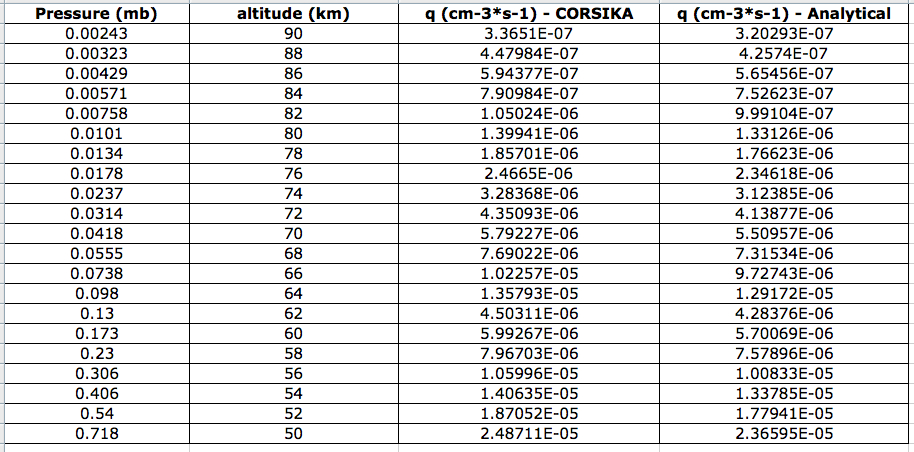}
\caption{Comparison of analytical approximation vs CORSIKA data
\newline for a thin target [21].}
\end{table}

\end{document}